# The Foundations of Genodynamics: The Development of Metrics for Genomic-Environmental Interactions

James Lindesay [1], Tshela E Mason [2], William Hercules[1,3], and Georgia M Dunston [2, 4,*]

[1] Computational Physics Laboratory, Howard University, Washington, DC, 20059, U.S.
E-mail: jlindesay@howard.edu (J.L.)

[2] National Human Genome Center, Howard University, Washington, DC, 20060, U.S.
E-mail: tmason@howard.edu (T.E.M.); gdunston@howard.edu (G.M.D.)

[3] Department of Physics, Morgan State University, Baltimore, MD, 21251, U.S.
E-mail: wmhercules@yahoo.com (W.H.)

[4] Department of Microbiology, Howard University, Washington, DC, 20060, U.S.
E-mail: gdunston@howard.edu (G.M.D.)

* Author to whom correspondence should be addressed; E-mail: gdunston@howard.edu; Tel: 202-806-7372

**Abstract:**
Single nucleotide polymorphisms (SNPs) represent an important type of dynamic sites within the human genome. These common variants often locally correlate into more complex multi-SNP haploblocks that are maintained throughout generations in a stable population. The information encoded in the structure of common SNPs and SNP haploblock variation can be characterized through a normalized information content (NIC) metric. Such an intrinsic measure allows disparate regions of individual genomes and the genomes of various populations to be quantitatively compared in a meaningful way.

Using our defined measures of genomic information, the interplay of maintained statistical variations due to the environmental baths within which stable populations exist can be interrogated. We develop the analogous "thermodynamics" characterizing the state variables for genomic populations that are stable under stochastic environmental stresses. Since living systems have not been found to develop in the absence of environmental influences, we focus on describing the analogous genomic *free energy* measures in this development.

The intensive parameter describing how an environment drives genomic diversity is found to depend inversely upon the NIC of the genome of a stable population within that environment. Once this environmental potential has been determined from the whole genome of a population, additive state variables can be directly related to the probabilities of the occurrence of given viable SNP based units (alleles) within that genome. This formulation allows the determination of both population averaged state variables as well as the genomic energies of *individual* alleles and their combinations. The determination of individual allelic potentials then should allow the parameterization of specific environmental influences upon shared alleles across populations in varying environments.

I. **Introduction**

**What is genodynamics?** We have introduced the term *"genodynamics"* to describe our *translational genomics research program* that is focused on exploring nucleotide structure-function relationships of common sequence variation and population genetics, grounded in first principles of thermodynamics and statistical physics (Lindesay et al, 2012). Using genodynamics, we study the informatics of single nucleotide polymorphisms (SNPs) as dynamic sites in the genome. Viewing structural configurations of SNPs as complex dynamical systems, we earlier utilized the normalized information content (NIC) as a biophysical metric for interrogating the information content (IC) present in SNP haploblocks. The NIC metric, derived from Boltzman's canonical ensemble and used in information theory, facilitates *translation* of biochemical DNA sequence variation into a biophysical metric for examining 'genomic-environmental interactions' at the nucleotide level. From this biophysical vantage point, the genome is perceived as a dynamic information system defined by patterns of SNP and SNP haploblock variation that correlate with genomic energy units (GEUs). The quantification of structural configurations of DNA sequence variation displayed in SNPs and SNP haploblocks into GEUs provides an additional biophysical metric for interrogating and translating the biology of common sequence variation. In this paper, we demonstrate that the information content of SNPs and SNP haploblocks correlate with genomic energy units that can be used to probe the structure-function relationship of genome organization. Using publically available bioinformatics tools, we have explored the functional significance of SNP haploblock variation in relationship to GEUs and evolutionarily conserved gene systems. Our findings demonstrate the utility of genodynamic metrics for exploring the biophysical foundation of sequence variation in translating the biology of genomic polymorphisms.

II. **Genodynamics and Energetics of the Human Genome**

The human genome consists of 3 billion nucleotides, most of which are fixed alleles. A significant number of sites (about 0.1%) non-randomly distributed across the human genome consist of single nucleotide polymorphisms (SNPs)(International Human Genome Sequencing Consortium, 2001). SNPs are (usually) bi-allelic dynamic sites on the human genome whose allelic distribution reflects the homeostasis of a population within a given environment. Here, environment will refer not only to geographic or geophysical parameters, but also to the complete interface of the population to biologic and evolutionary stresses. The viability of a population is directly reflected within those information measures that are maintained throughout generations.

In developing metrics for the interaction of the human genome with its environment, the genomic environment serves as a stochastic bath driving variations within a locally viable population. Since the SNPs are highly interacting dynamic sites, they are often very highly correlated into SNP haploblocks that are maintained with fixed frequencies within a given stable population. Combinations of SNPs that are very highly correlated within a population are said to be in *linkage disequilibrium*, indicating that only certain combinations of the alleles that make up the haploblock are linked in a manner that is generationally maintained. Population groups can therefore be *defined* by the maintained order and diversity of the genome as a functional whole with its environment.

One should also note that certain SNP allelic combinations *never* appear within the viable SNP haploblocks, indicating that the dynamically independent statistical micro-states are those combinations of SNP alleles describing haplotypes within SNP haploblocks, in addition to those associated with SNP sites that are not in linkage disequilibrium with any other SNPs. The linkage of several SNPs as conserved units that are passed between generations represents a type of statistical phase transition in forming complex dynamic units for a population within a given environment. The set of all SNP correlated variants in the genome encodes whatever information is maintained within a given population-environmental system. It is therefore very useful to develop information metrics that can quantify viable variation in the human genome.

**Entropy and information**

Information can be quantified in terms of the maintained order of a given system. In the physical sciences, the concept of *entropy* quantifies the dis-order of a physical system (Susskind, 2005). Therefore, entropy can serve as an additive measure of genodynamic variation within a population. This is done by taking the logarithm of multiplicative independent probabilities $p_h$, which define the *surprisals* $\log_2 p_h$. The specific (or per capita) entropy of a SNP haploblock consisting of a set of strongly dependent bi-allelic SNPs is taken to be the statistical average of this additive measure

$$s^{(H)} \equiv -\sum_h^{2^{n^{(H)}}} p_h^{(H)} \log_2 p_h^{(H)}, \qquad \text{(Eqn. 1)}$$

where $n^{(H)}$ is the number of bi-allelic SNP locations in haploblock H, and $p_h^{(H)}$ represents the probability (frequency) that haplotype $h$ occurs in the population. This measure of maintained (dis)order takes the value of zero for a completely homogeneous population with only one haplotype (since for $p_h^{(H)} = 1$, $\log_2 p_h^{(H)} = 0$), while it takes the value $s_{\max}^{(H)} = n^{(H)}$ for a completely stochastic distribution of all SNP alleles with all mathematically possible SNP haplotypes occurring with equal likelihood $p_h^{(H)} = \left(\frac{1}{2}\right)^{n^{(H)}}$.

For bi-allelic SNPs that are not in linkage disequilibrium, there are only 2 possible states at that location. Therefore, the entropy of the SNP location *(S)* takes the form

$$s^{(S)} \equiv -\sum_{a=1}^{2} p_a^{(S)} \log_2 p_a^{(S)}, \qquad \text{(Eqn. 2)}$$

where $p_a^{(S)}$ represents the probability (frequency) that allele $a$ occurs in the population. As defined here, the entropy has no dimensional units. The total specific entropy of the genome in the specified environment is given by the sum over all genetically viable blocks, including correlated SNPs in the haploblocks, along with individual SNPs between the haploblocks that are not in linkage disequilibrium,

$$s_{Genome} = \sum_H s^{(H)} + \sum_S s^{(S)}. \qquad \text{(Eqn. 3)}$$

This insures that *all* dynamic SNP degrees of freedom are included in calculating the genomic entropy. Because this entropy measure is additive, it also quantifies the entropy within any region of the genome. The overall entropy of a population distribution $S_{Population}$ is proportional to the size of the population $N_{Population}$, i.e $S_{Population} = N_{Population}\, s_{Genome}$, making entropy an extensive state variable.

Since entropy is a measure of the disorder of a distribution, a system with maximum *disorder* is one of maximum entropy. In contrast, the *information content* (IC) of a distribution is measured by the degree of *order* that the distribution has relative to a completely disordered one, i.e., the difference between the entropy of the distribution from that of a completely disordered distribution; $IC = S_{max} - S$. Such an information measure is likewise additive due to the additive nature of the entropy (Lindesay, 2013).

In our previous work (Lindesay, 2012), a normalized information metric was developed as a means of comparison of the information contained within specific regions of the genome, as well as between various populations. This normalized information content (NIC) value ranges between 0 and 1, where a value of zero indicates a completely random allelic distribution, while a value of unity represents a homogeneous allelic distribution without variation. The NIC for a given SNP haploblock (H) is defined by

$$NIC^{(H)} \equiv \frac{s_{max}^{(H)} - s^{(H)}}{s_{max}^{(H)}} = \frac{n^{(H)} - s^{(H)}}{n^{(H)}}. \qquad \text{(Eqn. 4)}$$

One should note that unlike the information content, NIC is *not* an additive measure for multi-SNP haploblocks. The information measure for the whole genome in an environment must be calculated using the total number of SNP locations in the genome, as well as the total specific entropy of the genome.

## Statistical energetics

The statistical "genomic energy" of a population in a given environment is expected to be an additive (extensive) state variable that depends upon the entropy, the populations of various allelic constituencies, and possibly the "genomic volume" of the environment, if population pressures have a significant effect upon the environment. This functional dependence of the contribution of haploblock H to this average genomic energy $U$ can be expressed using the differential expression

(Eqn. 5)

$$dU^{(H)} \equiv T_E \, dS^{(H)} + \sum_h m_h^{(H)} dN_h^{(H)} - \Pi_E^{(H)} dV^{(H)},$$

where $T_E$ represents an *environmental potential* (which is conjugate to the entropy state variable), $m_h^{(H)}$ represents the *allelic potential* of haplotype $h$ in SNP haploblock H, $N_h^{(H)} = p_h^{(H)} N_{Population}$ represents the population of haplotype $h$, and $\Pi_E^{(H)}$ represents any "partial pressure" by the haploblock on the environment that would result in contributing expansion of the genomic "volume" $V^{(H)}$. In all subsequent expressions, any genomic effects that would modify the genomic volume will be neglected $\Pi_E^{(H)} dV^{(H)} = 0$.

As is the case for thermodynamics and statistical physics, it is quite convenient to define an additive *free energy* state variable that is most naturally expressed as a function of the potential of the environmental bath $T_E$ and the populations, through the Legendre transformation

$$F^{(H)} \equiv U^{(H)} - T_E S^{(H)} \quad , \quad dF^{(H)} = -S^{(H)} dT_E + \sum_h m_h^{(H)} dN_h^{(H)}. \tag{Eqn. 6}$$

A focus upon the free energy as the fundamental dynamic state variable has the advantage of inherently including environmental-genomic interchanges as necessary considerations in describing the dynamics. It is a particularly convenient parameter for describing dynamics in a fixed environmental bath for which $dT_E=0$. As one recognizes that living cells have evolved their cellular functions within the warm, wet physiologic environment, one can safely conclude that a homeostatic living population distribution has evolved directly in association with the ecosystem within which it is being characterized. Thus, we assert that the evolution of living populations cannot be separated from their interchanges with the environment. In a statistical environment that is stochastically varying, it is the genomic free energy rather than the genomic energy that is minimized. The genomic free energy is a state variable that balances between conservation and variation of SNP haplotypes within an environment.

For the genome, only the site locations and bi-allelic nature of the specific SNPs are conserved parameters. However, the biological function associated with a given SNP location can be drastically altered by the allelic content of other SNPs within its proximity. In addition, phase transitions involving the stability of SNP haploblock structures are common between differing populations, resulting in non-conservation of the number and SNP composition of the haploblocks. This is in marked contrast to the standard micro-units in statistical physics, whose universal energy states are only weakly dependent upon the environment, and have well defined conservation properties with regards to the creation of new states (or changing dynamic degrees of freedom). Therefore, rather than seeking universal energy measures that are independent of the genomic environment, the emphasis here will be placed upon establishing convenient genomic measures of the dynamics that are inseparably coupled with environmental parameters. Since the allelic potentials, given by $m_h^{(H)} = \left( \dfrac{\partial F^{(H)}}{\partial N_h^{(H)}} \right)_{T_E}$, are the parameters in the environmental bath that dynamically couple to the SNP haplotype unit $h$, the formulation will be developed in a manner that most directly interprets these genomic energy measures.

Using the differential form for the genomic free energy $dF^{(H)} = -S^{(H)}dT_E + \sum_h m_h^{(H)} dN_h^{(H)}$ from (Eqn. 6), we can include the expression of the population with haplotype $h$ given by $N_h^{(H)} = p_h^{(H)} N_{Population}$ through expanding the differential $dN_h^{(H)} = dp_h^{(H)} N_{Population} + p_h^{(H)} dN_{Population}$. Re-writing the variation of the genomic free energy in terms of the population then gives

$$dF^{(H)} = \left(-s^{(H)} dT_E + \sum_h m_h^{(H)} dp_h^{(H)}\right) N_{Population} + \left(\sum_h m_h^{(H)} p_h^{(H)}\right) dN_{Population}. \quad \text{(Eqn. 7)}$$

## Population Stability

Values for all of the additive genomic state variables can be likewise assigned to those SNPs that are not in linkage disequilibrium by simply replacing the particular haploblock index *(H)* in any of the previous formulas with the SNP location *(S)*. The total genomic free energy will be a sum over all SNP haploblocks and non-linked SNPs given by

$$F_{Genome} = \sum_H F^{(H)} + \sum_S F^{(S)}. \quad \text{(Eqn. 8)}$$

We will next examine the condition that a stable population is defined by the genomic data. Our condition will require that the genomic free energy be a minimum under changes in the population within the static environment when the population is stable, i.e., $\left(\frac{\partial F_{Genome}}{\partial N_{Population}}\right) = 0$. From (Eqn. 7) for the genomic free energy in terms of block potentials and SNP potentials, holding the environmental potential and frequencies fixed, the population is seen to be *stable* if the overall genomic free energy satisfies

$$\left(\frac{\partial F_{Genome}}{\partial N_{Population}}\right)_{T_E, p_h} = 0 = \sum_H \left(\sum_h m_h^{(H)} p_h^{(H)}\right) + \sum_S \left(\sum_a m_a^{(S)} p_a^{(S)}\right) = \sum_H \langle m^{(H)} \rangle + \sum_S \langle m^{(S)} \rangle \equiv m_{Genome}.$$

(Eqn. 9)

Our population stability condition incorporates Hardy-Weinberg equilibrium (Hardy, 1908; Weinberg, 1908) in population genetics. Hardy-Weinberg equilibrium asserts that in order for the genomic distribution to meaningfully represent a stable population, the various frequencies of haplotypes and alleles should be mathematically stable within the distribution. Since the frequencies directly determine the block and SNP potentials, a requirement that these environmentally dependent potentials remain fixed and sum to zero satisfies Hardy-Weinberg equilibrium. Such stable populations maintain the distribution of SNPs throughout the generations within the given environment. The average allelic potential within a SNP haploblock $\sum_h m_h^{(H)} p_h^{(H)} = \langle m^{(H)} \rangle$ will be referred to as the *block potential* for haploblock *(H)*, while the average allelic potential at a non-linked SNP location $\sum_a m_a^{(S)} p_a^{(S)} = \langle m^{(S)} \rangle$ will be referred to as the *SNP potential* for location *(S)*. The genomic average allelic potential $m_{Genome}$, which is the sum over all block potentials and SNP potentials, is seen to vanish if the population does not increase or decrease. This means that a stable population is balanced with regards to its overall sum over allelic potentials, $m_{Genome}=0$. The genomic free energy is lowered by a population with negative overall genomic potential $m_{Genome}<0$ if its size increases, while if $m_{Genome}>0$ the genomic free energy is lowered if the population decreases.

As is the case in thermodynamics, the allelic potentials $m_h^{(H)}$ are expected to scale relative to the environmental parameter $T_E$, and allelic potential differences should directly reflect in the ratio of the frequencies of occurrence of those haplotypes within the population. A functional form that has these properties is given by

$$\frac{m_{h2}^{(H)} - m_{h1}^{(H)}}{T_E} = -\log_2 \frac{p_{h2}^{(H)}}{p_{h1}^{(H)}} . \qquad \text{(Eqn. 10)}$$

A *genomic energy unit* (GEU) $\tilde{m}$ will be defined as the unique allelic potential that will insure that a single (bi-allelic) SNP will be in its state of highest variation $\tilde{p} = \frac{1}{2}$ within the given species. Similarly, a haploblock with $n^{(H)}$ SNPs in its state of highest variation with all mathematically possible haplotypes occurring with frequencies $p_h^{(H)} = \left(\frac{1}{2}\right)^{n^{(H)}}$ will have an allelic potential of $n^{(H)} \tilde{m}$. The unit $\tilde{m}$ will be universal across all populations of a given species. Solving the previous equation, the allelic potential of the haplotype $h$ or allele $a$ in an environmental bath characterized by environmental potential $T_E$ can be expressed as

$$\begin{aligned} m_h^{(H)} &= (\tilde{m} - T_E) n^{(H)} - T_E \log_2 p_h^{(H)} \\ m_a^{(S)} &= (\tilde{m} - T_E) - T_E \log_2 p_a^{(S)} \end{aligned}, \qquad \text{(Eqn. 11)}$$

where the allelic potential for a single non-linked SNP location *(S)* has $n^{(S)} = 1$. Using our identifications, a lower allelic potential is then associated with a higher conservation of the SNP haplotype within the population, as high entropy is associated with large variation within the population. The ability to assign a well defined genomic energy measure for an *individual* haplotype once the environmental potential $T_E$ is known allows this formulation to establish biophysical measures beyond statistical statements about the population as a whole. Using this measure, the dependence upon environment of individual SNPs that are not in linkage disequilibrium, as well as SNP haploblocks that are shared between populations can be established by direct measurements.

The value of the allelic potential $m_a^{(S)}$ that fixes a single non-linked SNP location *(S)* into a given allele ($p_a^{(S)} \to 1$) will be defined to be the *fixing potential* in the given environment. If the allele has this potential, it is homogeneous throughout the population. From (Eqn. 11), this value is directly related to the environmental potential through

$$m_{Fixing} = \tilde{m} - T_E . \qquad \text{(Eqn. 12)}$$

Thus, the allelic potential of any single SNP location cannot be determined to be less than the fixing potential through measurements in a single environment.

The population stability condition $m_{Genome} = \sum_H \langle m^{(H)} \rangle + \sum_S \langle m^{(S)} \rangle = 0$ can be used to determine the environmental potential. By substituting the forms of the allelic potentials $m_h^{(H)}$ and $m_a^{(S)}$ expressed in terms of the probabilities into the population stability condition, an explicit expression of the environmental potential can be obtained:

$$T_E = \frac{\tilde{m} \, n_{SNPs}}{n_{SNPs} - s_{Genome}} = \frac{\tilde{m}}{NIC_{Genome}}, \quad (\text{Eqn. 13})$$

where $n_{SNPs} \equiv \sum_H n^{(H)} + \sum_S n^{(S)}$ is the total number of SNP locations on the genome. The average allelic potential for a given SNP haploblock, which has been defined as the block potential of that haploblock, then satisfies

$$\langle m^{(H)} \rangle = \left(1 - \frac{NIC^{(H)}}{NIC_{Genome}}\right) n^{(H)} \tilde{m}. \quad (\text{Eqn. 14})$$

This result has been obtained by simply taking the statistical average of the allelic potentials in (Eqn. 11), and substituting the expression for the environmental potential in terms of the genomic normalized information content (NIC).

These measures of genomic potentials have several convenient features:
- The environmental potential $T_E$ is inversely proportional to the information content (IC) of the whole genome. Low IC results from a high environmental potential, while a completely conserved genome has the lowest possible environmental potential, which with our definition has the value of one ***genomic energy unit*** $\tilde{m}$ =1 **GEU**. A population with a completely disordered genomic distribution would inhabit an environment with infinite environmental potential.
- SNP haploblocks that are highly conserved relative to the whole genome will have negative block potentials, while those that are highly varying will have positive block potentials. The block potentials typically lie within the range specified by $n^{(H)} m_{Fixing} \leq \langle m^{(H)} \rangle \leq n^{(H)} \tilde{m}$ (although the lower bound is not rigorously required).
- The number of highly correlated SNPs within the haploblock $n^{(H)}$ amplifies SNP haploblock allelic potentials.

One should note that while the environmental potential $T_E$, the block potentials $\langle m^{(H)} \rangle$, and the SNP potentials $\langle m^{(S)} \rangle$ can only be defined for a population as a whole, the individual allelic potentials $m_h^{(H)}$ and $m_a^{(S)}$ define an overall allelic potential for each individual in the population:

$$m_{Individual} = \sum_H m_h^{(H)} + \sum_S m_a^{(S)}, \quad (\text{Eqn. 15})$$

where the SNP haplotypes $h$ and alleles $a$ are unique to the individual. An individual's overall allelic potential is not a universal parameter, but rather depends strongly upon the environment. Thus, the overall allelic potential of an individual is not a fixed microphysical genomic energy state, in contrast to the energetics of particles in statistical physics. An environment within which an individual haplotype or allele has a negative allelic potential tends to conserve that characteristic, while an individual haplotype or allele that has a positive allelic potential provides diversity and viable genomic variation within that environment. The value of the allelic potential gives a direct measure of the dynamic (un) favorability of a haplotype as a function of the environment.

## III. Spectrum of block potentials

To demonstrate the usefulness of the previously defined genomic state variables, the parameters will be calculated using genomic data for stable populations. The populations that have been chosen for comparisons will be the Yoruba population of West Africa (YRI) and the central European population (CEU) (International HapMap Consortium, 2005). We have chosen representative large, medium and small chromosomes (1, 6, 11, 19, and 22) within the genome to examine the uniformity of the genomic potentials within the genome, and comparisons between populations. The analysis of data for the whole genome continues, and remains consistent with the results presented. Only the accuracy of the environmental potential $T_E$ is increased by including more chromosomes.

Our formulation requires that the SNP haploblock structure that codifies the linkage disequilibrium between local SNPs be established for a given population. For this purpose, we use Haploview, which is a software package in the public domain that is in general use. We constructed SNP haploblocks spanning chromosomes 1, 6, 11, 19, and 22 using the confidence interval algorithm developed by Gabriel et. al. in Haploview v 4.2 from HapMap phase III data. Haploview uses a two marker expectation-maximization algorithm with a partition-ligation approach which creates highly accurate population frequency estimates of the phased haplotypes based on the maximum-likelihood as determined from the unphased input (Barrett, 2005). Once the block structure of the population has been constructed, we have developed software that takes that data and calculates the genomic state variables.

The calculation of all genomic potentials requires that the environmental potential $T_E$ that bathes the genome of a given population be determined. This intensive parameter is directly related to the normalized information content (NIC) of the stable population in the given environment. For the chosen chromosomes, the distributions of the NIC values across the genomes of the two populations are demonstrated in Figure 1.

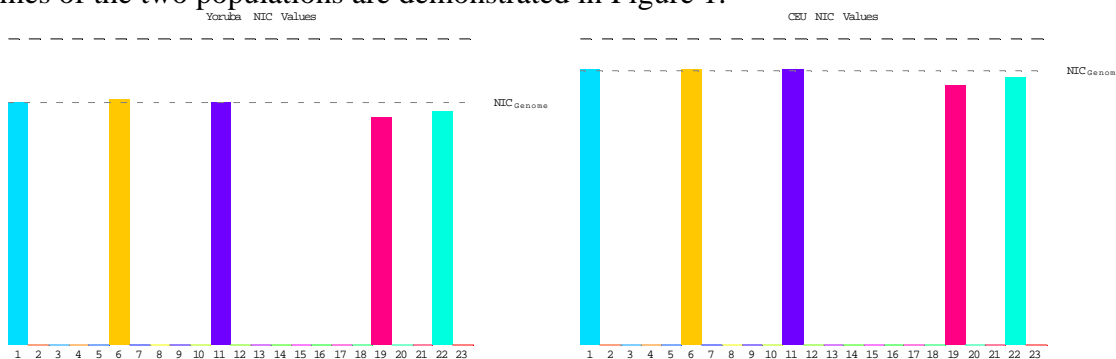

**Figure 1. Distribution of chromosomal NIC values. YRI left, CEU right.**

The uppermost dashed lines in the figure represent a NIC value of 1. Several features are apparent in this diagram. First, the YRI population has overall greater variation, while the CEU population exhibits more conservation, as quantified by its higher overall NIC. In both populations, it is clear that all chromosomes have NIC values within 10% of that of the overall regions of the genome that have been calculated. The larger chromosomes have NIC values that seem to be quite representative of that of each genome, while the smaller chromosomes seem to maintain slightly higher variation. Another feature of those chromosomes examined is that the relative distribution of conservation amongst the chromosomes seems to take the same shape between the two populations. Whether these features are fundamental properties of the human genome remains an unsettled question for further studies.

In order to demonstrate the usefulness of the genomic state variables, rather than overwhelm the reader with the abundance of data contained within all of the chromosomes that have been examined, the parameters will be demonstrated for chromosome 6 of both of the examined populations. It will be assumed that the environmental potential $T_E$ that would be calculated from the NIC of the whole genome does not differ significantly from that calculated using the five chromosomes included. This parameter takes the value $T_{E,(YRI)}$=1.26 GEUs for the YRI population, and $T_{E,(CEU)}$=1.12 GEUs for the CEU population.

Chromosome 6 has a total of $n_{SNPs}$=272,242 dynamic and non-dynamic SNP locations, of which 72% are within SNP haploblocks in the YRI population, and 87% within SNP haploblocks for the CEU population. By non-dynamic SNP locations, we mean that the alleles are fixed at those locations in this particular population. Across all chromosomes, those SNPs that are in linkage disequilibrium have higher NIC than those that are not. Generally, multi-SNP haploblocks favor stability and information conservation moreso than single SNP units. For the YRI population, the chromosomal NIC takes the value $NIC_6 \cong 0.80$, while for the CEU population, $NIC_6 \cong 0.90$. For the YRI population, the total (dimensionless) specific entropy of chromosome 6 has a value $s_6$=52,765, while in the CEU population it has the value $s_6$=26,633.

We next make comparisons of genomic energy measures between the individual SNP haploblocks within chromosome 6. We will develop a genomic energy spectrum by plotting the block potential of a haploblock in genomic energy units as a function of its location on the chromosome. The width of each bar on such a graph will indicate the locations that are in linkage disequilibrium defining a given SNP haploblock. Since the block potential is an average of the allelic potentials of the various haplotypes that make up the haploblock, such genomic energy spectra describe the population as a whole. The genomic energy spectra for chromosome 6 of the YRI and CEU populations are demonstrated in Figure 2.

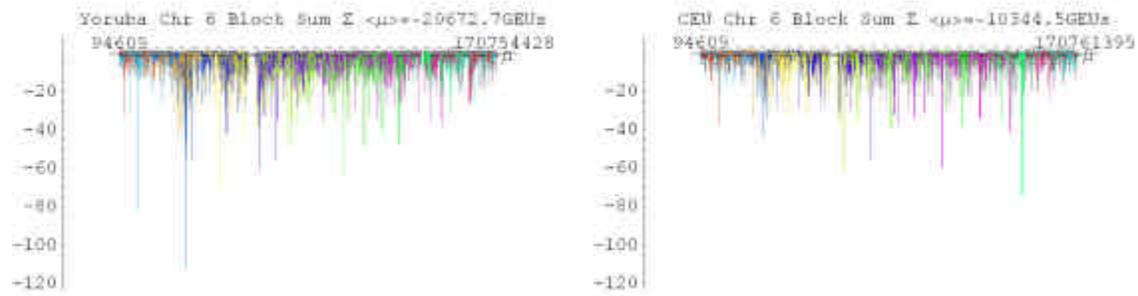

**Figure 2** Block spectra of SNP haploblocks in chromosome 6. YRI left, CEU right.

On the plots, the horizontal axis quantifies the location of the SNPs in each haploblock on the chromosome, while the vertical axis quantifies the block genomic energy in GEUs. The scale of the environmental potential $T_E$ is indicated by the uppermost dashed horizontal line, the average value for the allelic potential is indicated by the middle dashed horizontal line, and the fixing potential is indicated by the lowermost dashed horizontal line. The locations of the first and last SNP in linkage disequilibrium on the chromosome are explicitly displayed on the ends of the spectra. Since the overall chromosome has a very small value for the total block potential (from the population stability condition), it was initially expected that the spectrum would display an even distribution of positive and negative potentials. However, it is clear that although those haploblocks contributing positive block potential are uniformly distributed, those haploblocks contributing to negative block potentials are far fewer and more conserved. No highly varying SNP haploblock has a block potential significantly larger than the environmental potential $T_E$. Later we will focus in on the MHC region of this chromosome for a more detailed examination.

It is instructive to directly compare the informatics of chromosome 6 in the two populations. The NIC takes values between zero and one, where a value of zero indicates maximal variation in SNP haplotypes, while a value of one indicates complete sequence homogeneity of the population. We can plot the number of SNP haploblocks within given intervals of NIC as a measure of the proportion of haploblocks that maintain a specific degree of variation. Bar graphs of these proportions are demonstrated in Figure 3.

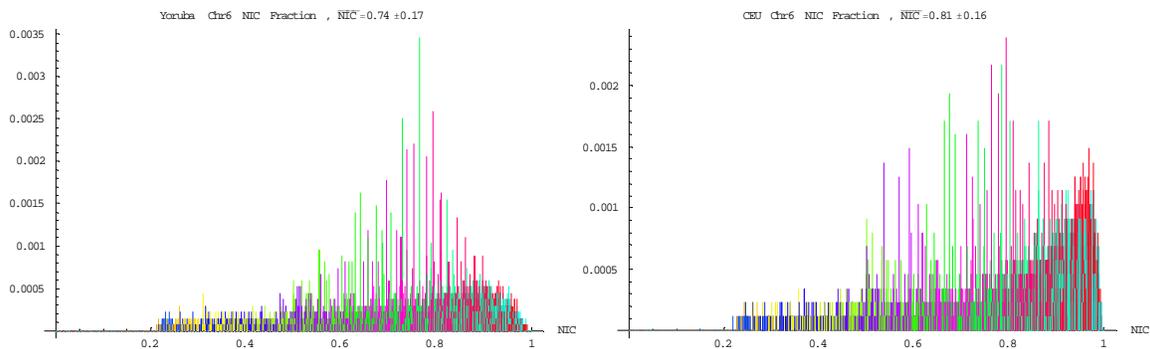

**Figure 3.** Normalized information content of haploblocks in chromosome 6. YRI left, CEU right.

In the plots, it is clear that the distribution for the CEU population is shifted relative to that of the YRI population, so that there are an increased number of SNP haploblocks with higher NIC. Haploblocks on chromosome 6 in the YRI population display a higher degree of variation than those in the CEU population overall.  Also, notice that there are *no* SNP haploblocks with a NIC value lower than ~0.2.  This indicates that many mathematically possible variations of alleles within the haploblocks are not viable within these stable human populations.

Prior to examining the genomic energy spectra in more detail, there is another interesting characteristic of the block potentials that was seen across all of the chromosomes examined.  This feature was discovered upon exploring the dependency of the block potentials upon the number of SNP locations within those haploblocks.  Although the block potential per SNP varied somewhat for haploblocks containing fewer than ~50 SNPs, the block potential per SNP for larger haploblocks was essentially constant within a given population, regardless of the chromosome examined.  For chromosome 6 of the two populations parameterized, the block potential as a function of the number of SNP locations in the haploblock is plotted in Figure 4.

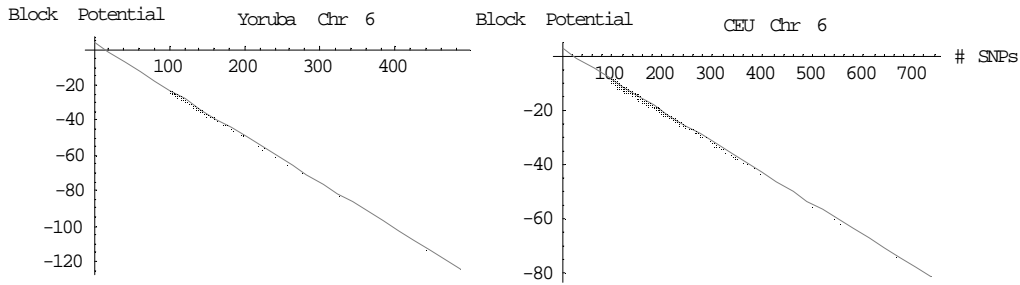

**Figure 4  Plots of the block potential as a function of the number of SNP locations for large haploblocks.  YRI left, CEU right.**

A set of good fits for the block potentials for large haploblocks are given by
$$m_{YRI}^{(H)} \cong (3.84 - 0.26 n^{(H)}) GEUs$$
$$m_{CEU}^{(H)} \cong (2.81 - 0.11 n^{(H)}) GEUs$$
where again, $n^{(H)}$ is the number of SNP locations in haploblock H.  These results indicate that those SNP haploblocks that are highly conserved are seen to have block potentials that approach the fixing potential $\mu_{\text{fixing}}$ for the given specific environment times the number of SNP locations in the block.

## The MHC region of chromosome 6

To further examine the biophysical interpretations of genomic energies, we will consider the block spectrum of SNP haploblocks in the MHC region for the populations under consideration.  This region is important in encoding immune response, and is expected to be particularly sensitive to environmental influences.  The block genomic energy spectra for the region from 26,195,488 to 33,389,967 on chromosome 6 for the YRI and CEU populations are demonstrated in Figure 5.

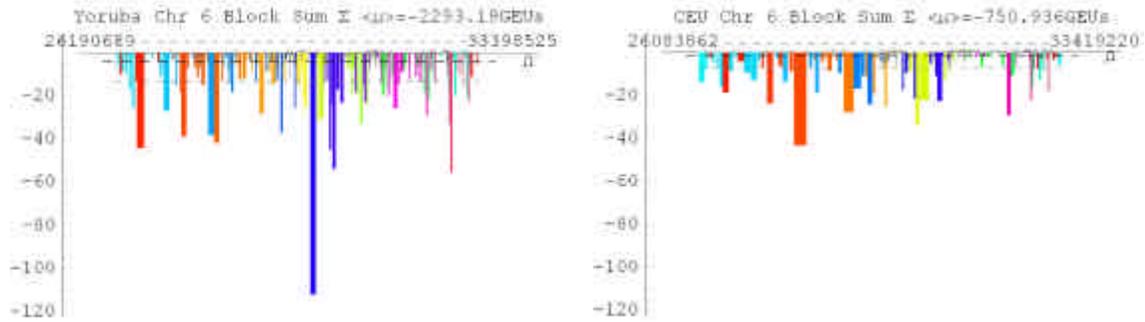

**Figure 5. Block spectra for SNP haploblocks of MHC region. YRI left, CEU right.**

One striking feature of comparison between the spectra is that despite the overall higher diversity of the YRI population as quantified in its considerably lower NIC when compared to the CEU population, within the MHC region the SNP haploblocks of the YRI population has genomic energies that are considerably more conserved $\sum_H m^{(H)} \cong -2293 GEUs$ compared to those of the CEU population $\sum_H m^{(H)} \cong -751 GEUs$. This is also apparent from the lower value for the average block potential demonstrated by the middle dashed lines on the plots. In particular, the most highly conserved haploblock that has been found within any of the chromosomes thus far examined is the block 3013 on chromosome 6 in this region of the YRI population. We will examine the biological functions within this block later.

In order to provide a direct comparison of the block potentials with their corresponding information contents, plots of the NIC vs. block location are illustrated in Figure 6.

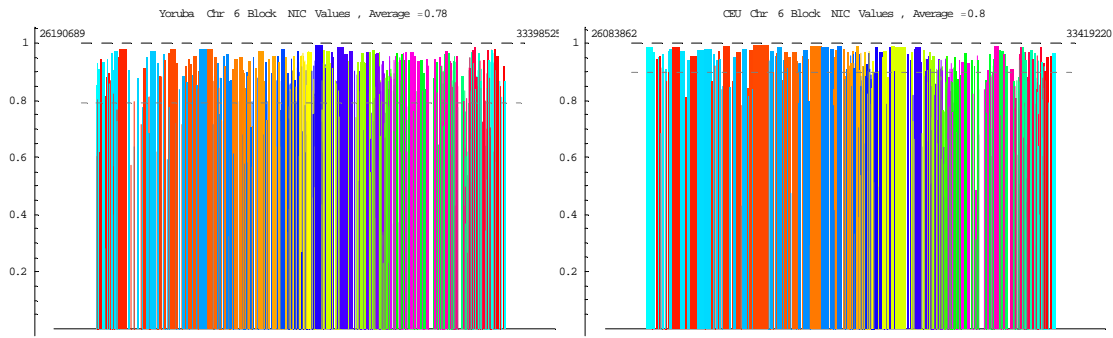

**Figure 6. NIC value vs. block location for haploblocks of MHC region. YRI left, CEU right.**

In the plots, the shadings and locations on the chromosomes have been maintained consistent with Figure 5. The upper dashed horizontal line represents the upper limit of NIC (unity), while the lower dashed lines represent the (approximated) whole genome NICs for the populations.

The distributions of values of the NICs of the SNP haploblocks in this region are compared between the examined populations in Figure 7.

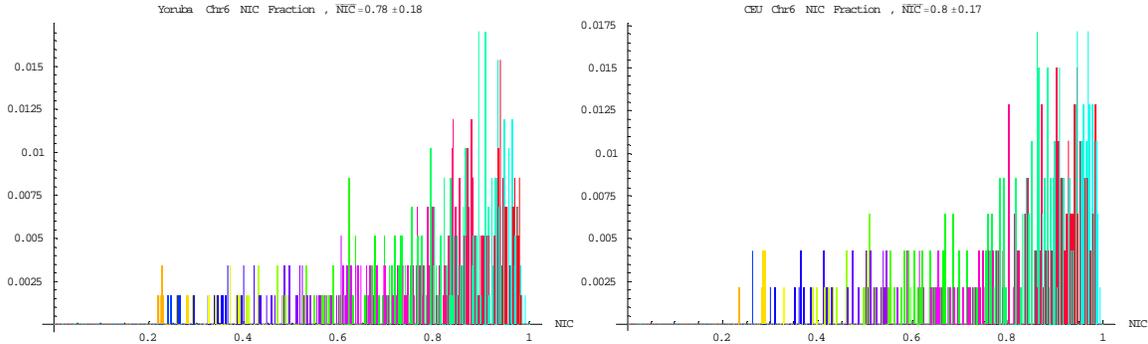
**Figure 7. NIC for SNP haploblocks of MHC region. YRI left, CEU right.**

This figure demonstrates a shift in the distribution of NIC towards higher values in the CEU population relative to its distribution in the YRI population. For both populations, the average NIC value for this region is higher than that of each overall genome. Our prior studies (Lindesay, 2012) demonstrated that those SNP haploblocks with high NIC tend to be associated with genic regions of the chromosome, while those with low NIC were associated with innate immune regulation and functions that require rapid response to environmental stresses.

We can also examine the genomic energy spectrum of those SNP locations that are not in linkage disequilibrium with any other SNP location. The SNP spectra for the populations examined are demonstrated in Figure 8.

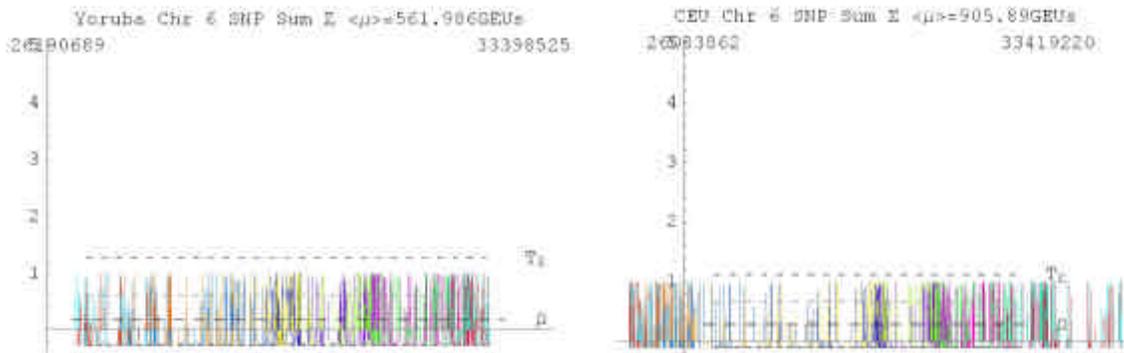
**Figure 8. SNP spectrum for non-linked SNPs of MHC region. YRI left, CEU right.**

The contribution of the SNP potentials in this region for the YRI population is $\sum_S m^{(S)} \cong +562 \, GEUs$ is also lower than that for the CEU population $\sum_S m^{(S)} \cong +906 \, GEUs$. It is clear from the positive contribution of these SNPs for both populations that these non-linked SNPs contribute to variation within the overall genomes in each environment. An additional feature of these spectra is of interest. The SNP potential for individual SNPs is constrained between the fixing potential and the environmental potential. Although the upper limit is defined through the definitions of the genodynamic parameters, the lower limit can in principle be lower than the fixing potential. If such is the case, its values for individual SNPs would have to be extrapolated using the dynamic dependencies of those SNPs upon environmental parameters derived from multi-population studies. A significant lowering of many of these potentials could have an impact on the calculation of the environmental potential. We are leaving such examinations to future studies.

The distributions of the NICs of the non-linked SNPs in this region are compared between the populations in Figure 9.

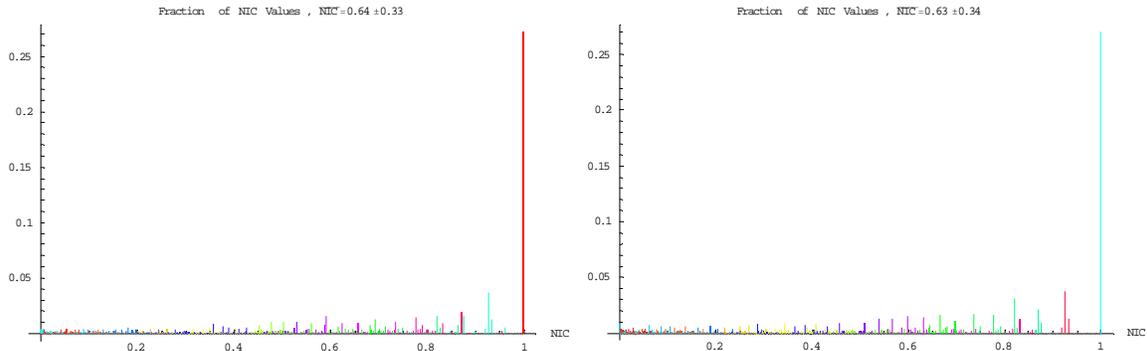
**Figure 9. NIC distribution for non-linked SNPs of MHC region. YRI left, CEU right.**

Both populations display a significant number (~25%) of highly conserved SNPs that are fixed within the given environments, but an otherwise broad distribution of SNPs over all other NIC values.

*Genomic energies of individual alleles*

As previously mentioned, once the environmental potential has been determined for a stable population distribution in a specific (stationary) environment, individual genomic energies can be determined for specific haplotypes and alleles within that environment based upon their relative frequencies. A haplotype in an environment that highly favors its conservation will have a significant negative allelic potential lowering the overall genomic free energy of that individual genome. In addition, the variation of individual allelic potentials amongst differing populations can in principle be used to parametrically track the *specific* environmental dependencies of those potentials. Although we will not explore detailed comparative features of individual alleles within environments in this paper, we will examine exemplar alleles in the MHC region of the YRI population to illustrate the calculation of individual allelic potentials.

We first illustrate the genomic energy of a SNP that is very nearly, but not quite completely, conserved. The SNP rs886420 at location 30987615 on YRI chromosome 6 is not in linkage disequilibrium with any other SNP in this population, and is polymorphic for alleles C and T. This nearly conserved SNP has a (very low) specific entropy of 0.04 and a (very high) NIC value of 0.96. Figure 10 shows the individual allelic potential for rs886420:

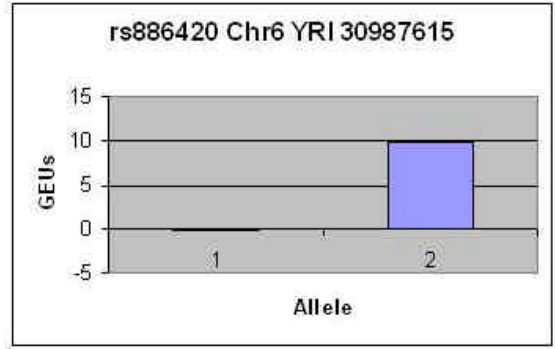

**Figure 10 The C and T allelic potentials.**

The contribution of the SNP location to the overall SNP potential is -0.2GEUs. In particular, the allelic potential of the dominant allele is quite near the fixing potential μ$_C$≅-0.26GEUs, while that of the nearly absent allele is quite high, ***m***$_T$≅+9.8GEUs. There are many other SNPs in the MHC region with SNP potentials near the fixing potential. A few other SNPs with allelic potentials indistinguishable from that of rs886420 are rs3130544, rs1265096 and rs9357117.

Another illustrative exemplar is a SNP location at which the alleles exhibit the maximum variation. The SNP rs2734335 at location 32001923 on YRI chromosome 6 is essentially stochastic, contributing a (very high) specific entropy of 0.999954 with highest variation as indicated by its NIC value 0.000046 ≈0. It is not in linkage disequilibrium with any other SNP in this population, and is polymorphic in alleles A and G. Its contribution to the overall SNP potential essentially *defines* the standard GEU for the human population, $\langle \boldsymbol{m}^{(S)} \rangle \cong 0.999942\, \tilde{\boldsymbol{m}} \approx 1 GEU$ as illustrated in Figure 11.

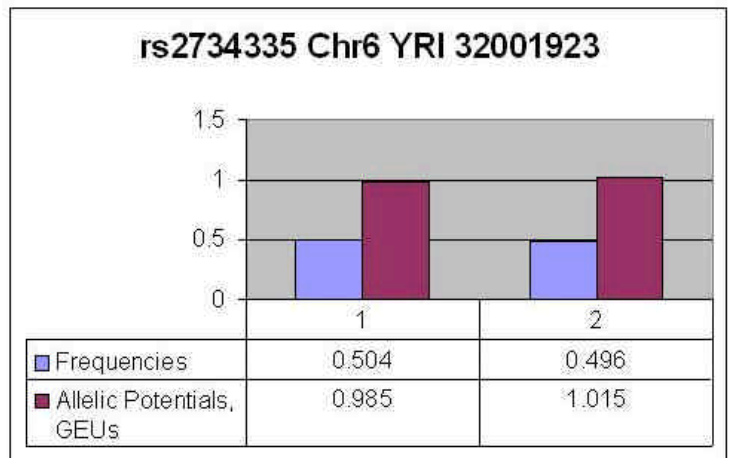

**Figure 11. The A and G allelic potentials and frequencies.**

This SNP represents alleles with the highest possible degree of variation in the given environment, and contributes the highest allelic potential that a single SNP can contribute to the overall genomic free energy.

As exemplars of non-linked SNP locations with *zero* SNP potential, consider the two locations (rs9276192 at 32807454) and (rs9461841 at 33189191) on chromosome 6 of the YRI population. At each of these locations, the allele G occurs with a probability of 0.97 in the population, while the allele A occurs with a probability of 0.033 in the population, resulting in specific entropy of 0.21 and a NIC value of 0.79 for each of these SNPs. These frequencies, as well as the resultant allelic potentials, are plotted in Figure 12.

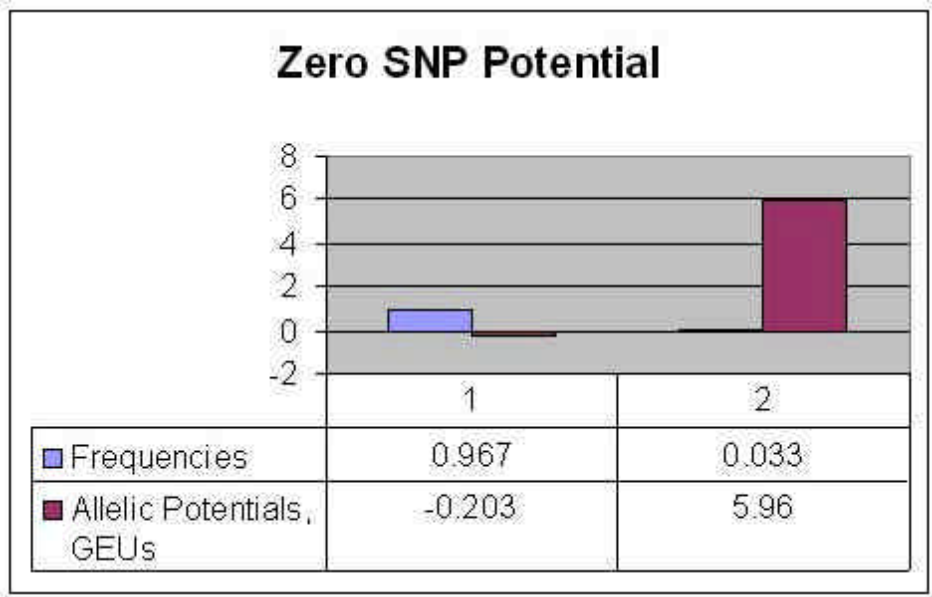

**Figure 12. The A and G allelic potentials and frequencies for SNPs with essentially zero SNP potential**

From this figure, it is clear that although the average SNP potential vanishes, the allelic potentials of the individual alleles do not. Such SNPs with zero SNP potential contribute no cost or benefit to the genomic free energy within the given environment.

We will next examine larger units of maintained information encoded in exemplar SNP haploblocks. A typical SNP haploblock within the YRI population is demonstrated by Block 2836 on chromosome 6, which has 7 SNP locations, (5 of which are dynamic in this population) contributing a specific entropy of 1.42, with a NIC value of 0.80, essentially the same as that for the whole 5 chromosome region of the YRI genome explored to this point. The allelic spectrum of this haploblock is demonstrated in Figure 13.

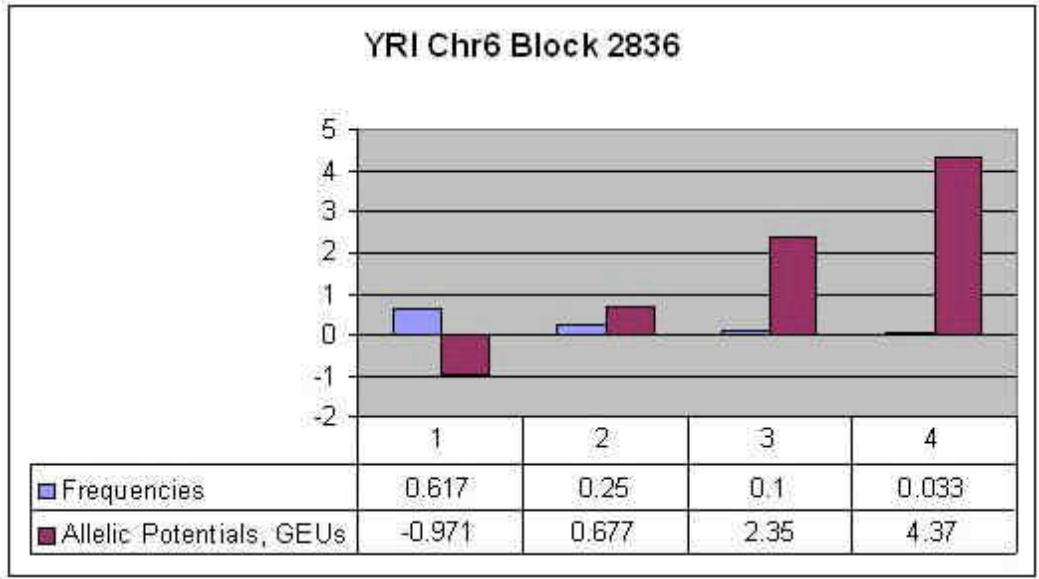

**Figure 13. Allelic potentials and frequencies for the haplotypes in Block 2836.**

Because the NIC of this haploblock is very nearly the same as that of all of the chromosomes so far examined, it contributes an essentially negligible block potential of -0.05GEUs to the overall genomic potential of the population, providing a multi-SNP example of the no-GEU-cost SNPs discussed in Figure 12.

*A highly conserved haploblock*

We will end by examining the very highly conserved block 3013 on chromosome 6 of the YRI population. This haploblock had the lowest block potential of any on the 10 chromosomes in the 2 populations examined thus far. Haploblock 3013 of the 13548 SNP haploblocks on chromosome 6 of the Yoruba population is located in the MHC region between locations 30,058,189 to 30,164,455. The specific entropy of this haploblock is 3.6 and its NIC is 0.992. The allelic spectrum of its various haplotypes in this environment is demonstrated in Figure 14.

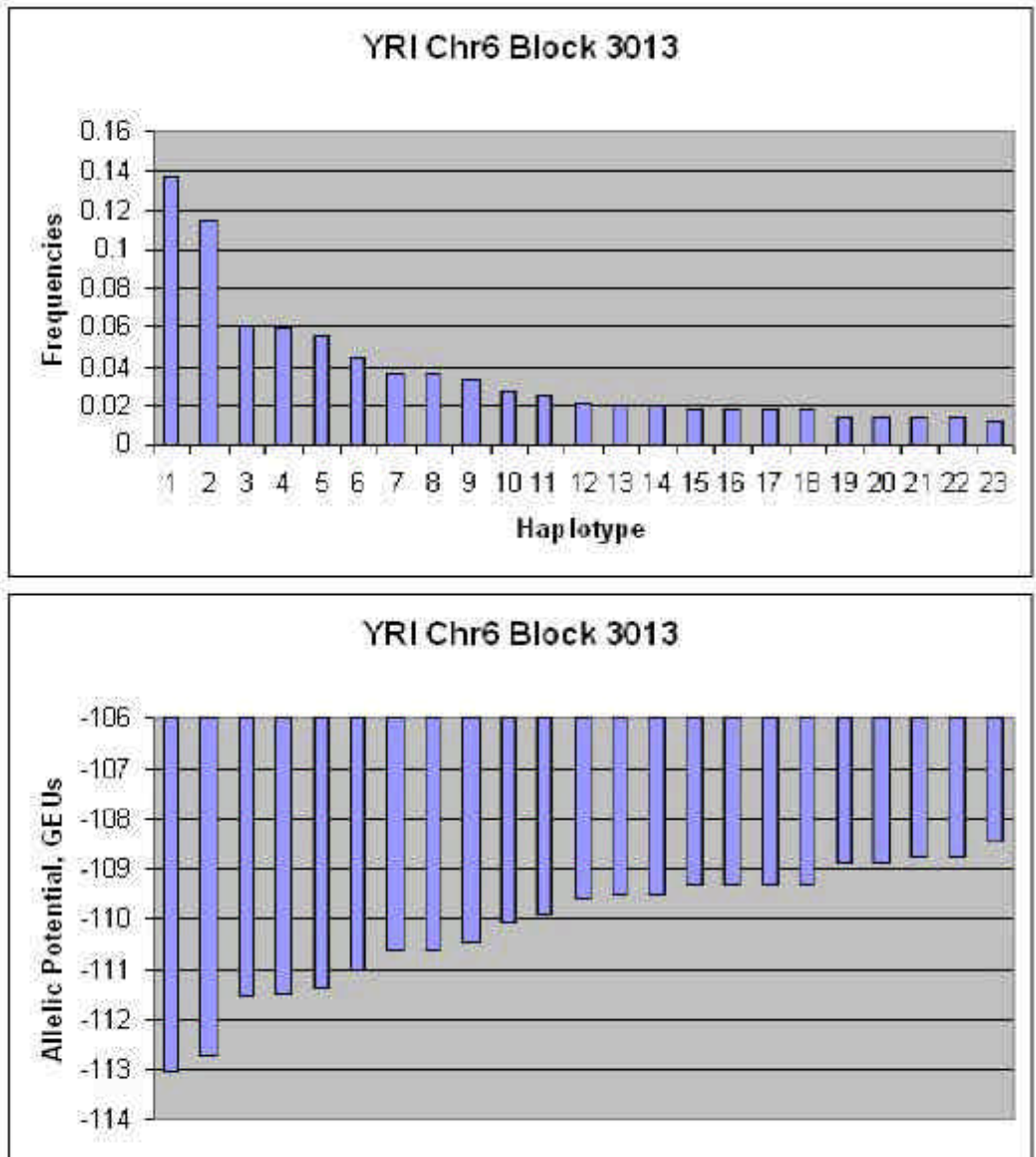

**Figure 14. Allelic potentials and frequencies of the haplotypes in Block 3013.**

Generally, for a highly conserved haploblock, the allelic potentials of the haplotypes are all energetically favorable. Block 3013, which has 441 SNP locations (226 dynamic SNPs in this population), contributes a highly favorable averaged allelic potential of -112 GEUs to the overall genomic distribution. Block 3013 has 253 SNPs in genes and 188 SNPs in non-genic regions. This block spans the following six genes: Zinc ribbon domain 1 (ZNRD1); ZNRD1-antisense RNA1 (ZNRD1-AS1); HLA complex group 8 (HCG8); Protein phosphatase 1 regulatory inhibitor subunit 11 (PPP1R11); Human leukocyte antigen J (HLA-J) and Ring finger protein 39 (RNF39). ZNRD1-AS1 is what is known as a *natural antisense transcript* (NAT). NATs are highly conserved regulatory mechanisms which span across species, from viruses (Scherbakov, 2000) to fungi (Havilio, 2005; David, 2006) to plants (Wang, 2005; Jen, 2005) to mammals (Chen, 2004; Yelin, 2003; Zhang, 2006; Kiyosawa, 2003; Wang, 2008; Li, 2008; Sun, 2006), suggesting that this mode of gene regulation is *evolutionarily* conserved. For comparison, we also examined the most highly conserved haploblock with a highly favorable allelic potential in the CEU population. Block 7016 is located on chromosome 6 as well between locations 145,851,676 to 146,351,676. This block has an allelic potential of -73.85 GEUs and contains 666 SNPs, with 353 of them being dynamic. The specific entropy of this block is 3.1 and its NIC value is 0.996. The block has 399 SNPs in genes and 267 SNPs in non-genic regions. Block 7016 spans the following four genes: Epilepsy, progressive myoclonus type 2, Laforin disease [laforin] (EPM2A); Uncharacterized protein (LOC100507557); SNF2 histone linker PHD RING helicase E3 ubiquitin protein ligase (SHPRH), and F-box protein 20 (FBXO30). Like ZNRD1-AS1, LOC100507557 is also a NAT. This means that the SNP haploblock with the lowest GEUs in both the YRI and the CEU populations contain a NAT. Thus, using genomic energy units (GEUs) as a biophysical metric in SNP haploblock analysis has provided insight on the inheritance of highly conserved NAT regulatory mechanisms encoded in the structure of common variants.

.

## IV. Conclusions

We have developed genomic energy measures for the human genome that relate the distribution of alleles within a stable population to state variables associated with the environment within which that population resides. The state variables defined by common variations utilize the entropy of the statistical distribution of alleles to establish normalized information measures for persistent dynamic units within arbitrary regions of the genome, as well as for the genome as a whole. The normalized information content (NIC) of the whole genome has been found to determine an overall environmental potential that is a state variable that parameterizes the extent to which the environment drives variation and diversity within the population. Once this environmental potential (which is canonically conjugate to the entropy) has been determined, the genomic energies of individual alleles and sets of alleles, as well as statistical average genomic energies for each persistent dynamic unit, can be directly calculated. We showed that the most highly conserved SNP haplotype regions correlate with *evolutionarily* conserved regulatory mechanisms, such as NATs.

The assignment of genomic energies to individual alleles within a given environment allows the parameterization of specific environmental influences upon shared alleles across populations in varying environments. We are examining simple allelic dependencies upon single environmental parameters for future presentation.

## Acknowledgments

The authors wish to convey appreciation for the continuing support of the National Human Genome Center, and the Computational Physics Lab, at Howard University. We also acknowledge Zahra Dawson for developing computer programs for calculating biophysical metrics from Haploview raw data.

# Bibliography

Barrett JC, Fry B, Maller J, Daly M.J. (2005) Haploview: analysis and visualization of LD and haplotype maps. Bioinformatics; 21: 263-265.

Chen J et al. (2004) Over 20% of human transcripts might form sense-antisense pairs. Nucleic Acids Res; 32(16): 4812-4820.

David L et al. (2006) A high-resolution map of transcription in the yeast genome. Proc Natl Acad Sci USA; 103(14):5320-5325.

Hardy GH. (1908) Mendelian proportions in a mixed population. Science; 28:49-50.

Havilio M et al. (2005) Evidence for abundant transcription of non-coding regions in the Saccharomyces cerevisiae genome. BMC Genomics; 6(6): 93.

International Human Genome Sequencing Consortium. (2001) Initial sequencing and analysis of the human genome. Nature 409:860-921.

International HapMap Consortium. (2005) A haplotype map of the human genome. Nature 437:1299-320

Jen CH et al. (2005) Natural antisense transcripts with coding capacity in Arabidopsis may have a regulatory role that is not kinked to double-stranded RNA degradation. Genome Biol; 6(6):R51.

Kiyosawa H, et al. (2003) Antisense transcripts with FANTOM2 clone set and their implications for gene regulation. Genome Res, 13(6B): 1324-1334.

Li JT et al. (2008) Trans-natural antisense transcripts including noncoding RNAs in 10 species: implications for expression regulation. Nucleic Acids Res; 35(15): 4833-4844.

Lindesay J (2013). **Foundations of Quantum Gravity.** Cambridge University Press, Cambridge, UK.

Lindesay J et al. (2012). A New Biophysical Metric for Interrogating the Information Content in Human Genome Sequence Variation: Proof of Concept. J Comput Biol Bioinform Res; 4(2): 15-22.

Scherbakov DV and Garber MB. (2000) Overlapping Genes in Bacterial and Phage Genomes. Molecular Biology, 34(4):485-495.

Sun M et al. (2006) Evidence for variation in abundance of antisense transcripts between multicellular animals but no relationship between antisense transcriptional organismic complexity. Genome Res 16(7):922-933.

Susskind L and Lindesay J (2005). **An Introduction to Black Holes, Information and the String Theory Revolution.** World Scientific Publishing Company, New Jersey, US.

Wang A, et al. (2008) Molecular characterization of the bovine chromodomain Y-like genes. Anim Genet; 39(3): 207-216.

Wang XJ, Gaasterland T and Chua NH. (2005) Genome-wide prediction and identification of cis-natural antisense transcripts in Arabidopsis thaliana. Genome Biol; 6(4):R30.

Weinberg W. (1908) Über den Nachweis der Vererbung beim Menschen. Jahresh. Ver. Vaterl. Naturkd. Württemb; 64**:** 369–382.

Yelin R, et al. (2003) Widespread occurrence of antisense transcription in the human genome. Nat Biotechol, 21(4):379-386.

Zhang Y et al. (2006) Genome-wide in silico identification and analysis of cis natural antisense transcripts (cis-NATs) in ten species. Nucleic Acids Res; 34(12): 3465-3475.

**Bioinformatic Tools utilized for this manuscript**: HapMap, dbSNP, GENE, Regulome dB, NATS dB, F-SNP, BioGPS.